\documentclass[11pt]{article}
\usepackage[utf8]{inputenc}
\usepackage{amsmath,setspace,geometry}
\usepackage{amsthm}
\usepackage{amsfonts}
\usepackage[shortlabels]{enumitem}
\usepackage{rotating}
\usepackage{pdflscape}
\usepackage{graphicx}
\usepackage{bbm}
\usepackage{comment}
\usepackage[dvipsnames]{xcolor}
\usepackage{hyperref}
\hypersetup{colorlinks=true, linkcolor= BrickRed, citecolor = BrickRed, filecolor = BrickRed, urlcolor = BrickRed, hypertexnames = true}
\usepackage[]{natbib} 
\bibpunct[:]{(}{)}{,}{a}{}{,}
\usepackage[english]{babel}
\usepackage{float}
\usepackage{caption}
\usepackage{subcaption}
\usepackage{booktabs}
\usepackage{pdfpages}
\usepackage{threeparttable}
\usepackage{lscape}
\usepackage{bm}
\usepackage{setspace}
\title{Nonparametric Estimation of Matching Efficiency and Elasticity in a Marriage Agency Platform: 2014--2025}
\author{Suguru Otani\thanks{\href{mailto:}{suguru.otani@e.u-tokyo.ac.jp}, Market Design Center, Department of Economics, The University of Tokyo\\
 I thank Fuhito Kojima, Chiappori, Pierre-Andr{\'e}, Kadachi Ye, Shangwen Li, Yusuke Ishihata, Chihiro Inoue for their valuable advice. I acknowledge Shuto Fukuda and Hirokazu Tsuchiya for sharing the data and their technical and institutional knowledge in the IBJ platform. This work was supported by JST ERATO Grant Number JPMJER2301, Japan.\\Declarations of interest: none } }
 
\date{
\today
}

\begin{document}

\maketitle

\begin{abstract}
    This paper examines monthly matching efficiency in the Japanese marriage market using novel data from IBJ, the country’s largest structured matching platform. Unlike administrative or dating app data, IBJ provides full search, dating, and matching logs based on verified profiles and confirmed engagements. Users are highly selected into serious marriage search via costly screening. Covering 3.3\% of national marriages annually, I estimate time-varying matching functions using a nonparametric approach. Matching efficiency triples between 2014 and 2025, with diverging elasticities indicating increasingly asymmetric responsiveness. Results show how digital intermediation reshapes partner search in modern marriage markets.
    \quad \\
    %100 words AER (now 100 words)
\end{abstract}

\section{Introduction}
Understanding how individuals find partners in modern marriage markets is central to the study of family economics \citep{chiappori2023mating}. As marriage rates decline and matching technologies proliferate, questions surrounding search efficiency, partner availability, and market design have grown increasingly salient. While theoretical models of two-sided search abound, empirical research has been hampered by data limitations: traditional sources such as administrative records or surveys capture only realized marriages, omitting the entire search and matching process that precedes them, as well as the presence of singles who are not seriously seeking partners. Consequently, little is known about the dynamics of partner search, how modern platforms shape matching outcomes, or how efficiently modern markets operate.
This study addresses these limitations by using novel data from IBJ, Japan's largest structured marriage matching platform, which accounts for approximately 3.3\% of all new marriages in Japan in 2024.\footnote{Approximately 50\% of engagements among IBJ users occur within the platform, while the remaining 50\% result from relationships formed outside the platform.} The dataset includes over 10,000 confirmed engagements annually and allows for behavioral analysis of the full matching process, including search, proposals, and formal engagements. In this note, we focus on documenting macro-level trends on the platform as a first step toward more detailed analysis, leaving in-depth investigations of user behavior and underlying mechanisms to future work.

There are four distinctive and advantageous features of the IBJ data compared to dating apps \citep{hitsch2010matching,ong2015income,bapna2016one,egebark2021brains,rios2023improving}, speed date events \citep{fisman2006gender,belot2013dating}, and government records \citep{chiappori2017partner}. First, the data contain rich and verified user demographics, which are linked to official records such as tax withholding documents and certified health checkup results. Second, users are highly selected into serious partnership search due to substantial screening and financial costs. Third, the dataset records full, time-stamped logs of search and interaction behavior linked to confirmed outcomes, and includes information on whether users were active or inactive in their marriage search at each point in time. Fourth, the data are observed at the level of individual user behavior logs, enabling high-frequency analysis.
Importantly, the IBJ data's richness and granularity, combined with its one-to-one matching structure, enable a direct analogy to labor market matching function estimation \citep{petrongolo2001looking}, allowing us to assess platform-level efficiency and responsiveness. 

Using this dataset, I estimate a matching function following the nonparametric approach of \citet{lange2020beyond} to evaluate the platform's performance.\footnote{Their approach is widely applied to labor market settings. See \cite{otani2024onthejob,otani2024nonparametric} and \cite{kanayama2024nonparametric}.} I find that matching efficiency increased threefold between 2014 and 2025, with gains beginning in 2017, coinciding with major platform consolidation. Engagement formation also became more responsive—though increasingly asymmetrically—to changes in user composition, as indicated by diverging elasticities on the male and female sides. These results underscore the role of digital intermediation in shaping modern family formation. Unlike traditional arrangements, platforms like IBJ provide structured, data-rich environments where frictions are minimized, outcomes are verified, and behavioral dynamics—potentially differing across user groups—are observable, enabling new empirical insights.

\section{Data}
I use confidential data from IBJ covering the period 2014--2025. In 2024, IBJ accounted for 3.3\% of all new-marriages in Japan, with over 10,000 engagements annually. While this share is smaller than administrative marriage registers, IBJ offers unique visibility into the pre-marriage process. The platform collects verified user data—age, income, education, and marital status—and tracks the full sequence of actions: search, proposals, messages, and dates. All records are time-stamped and linked to engagement outcomes.
IBJ users are highly selected into serious long-term partnership search. Entry requires screening and substantial upfront, monthly, and exit fees, effectively excluding casual users. In contrast to dating apps with self-reported profiles and open-ended interactions, IBJ is a closed, consultant-mediated two-sided platform where mutual consent leads to confirmed engagements. This structured setting makes the data well-suited for analyzing matching efficiency and responsiveness over time.

\begin{figure}[!ht]
  \begin{center}
  \subfloat[Female $F$, Male $M$, and tightness ($\frac{F}{M}$)]{\includegraphics[width = 0.33\textwidth]{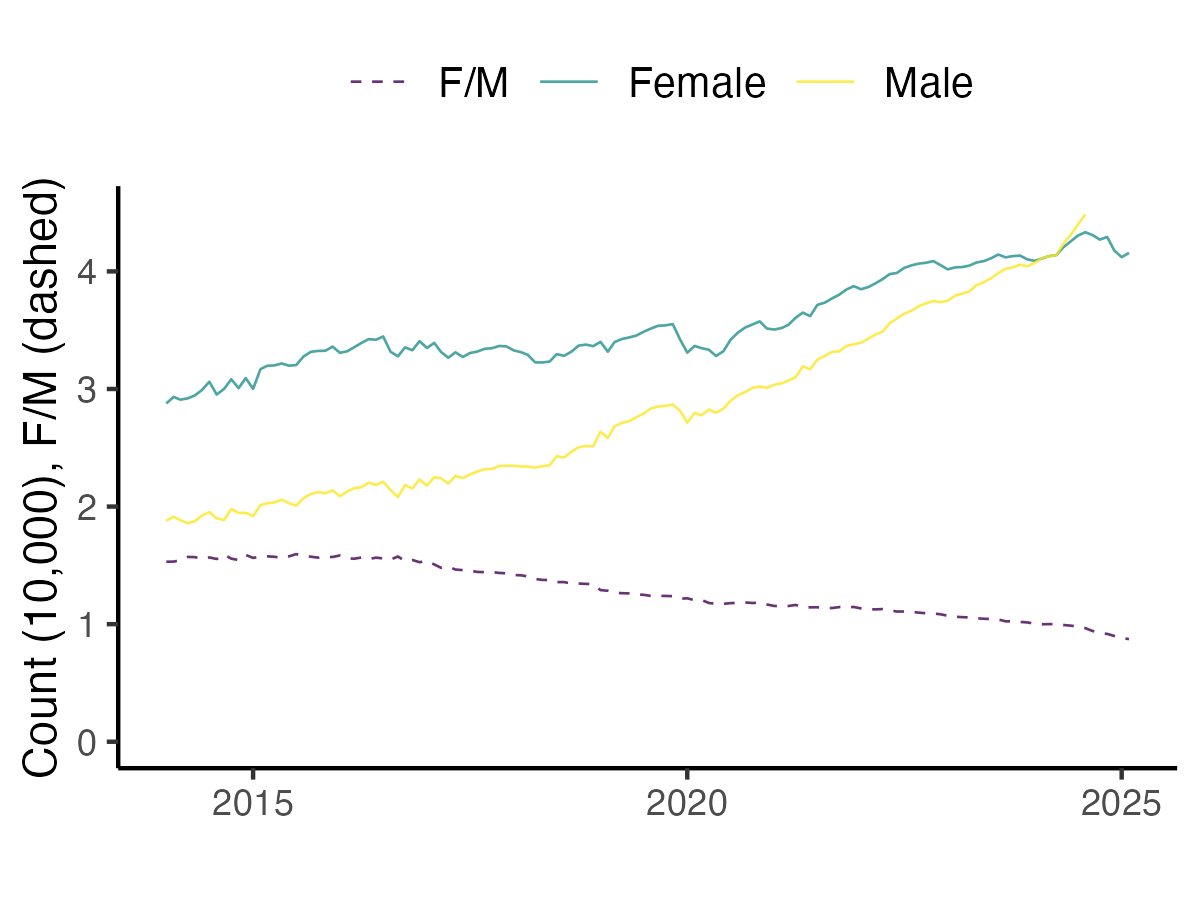}}
  \subfloat[(Within-platform) Engagement $E$]{\includegraphics[width = 0.33\textwidth]{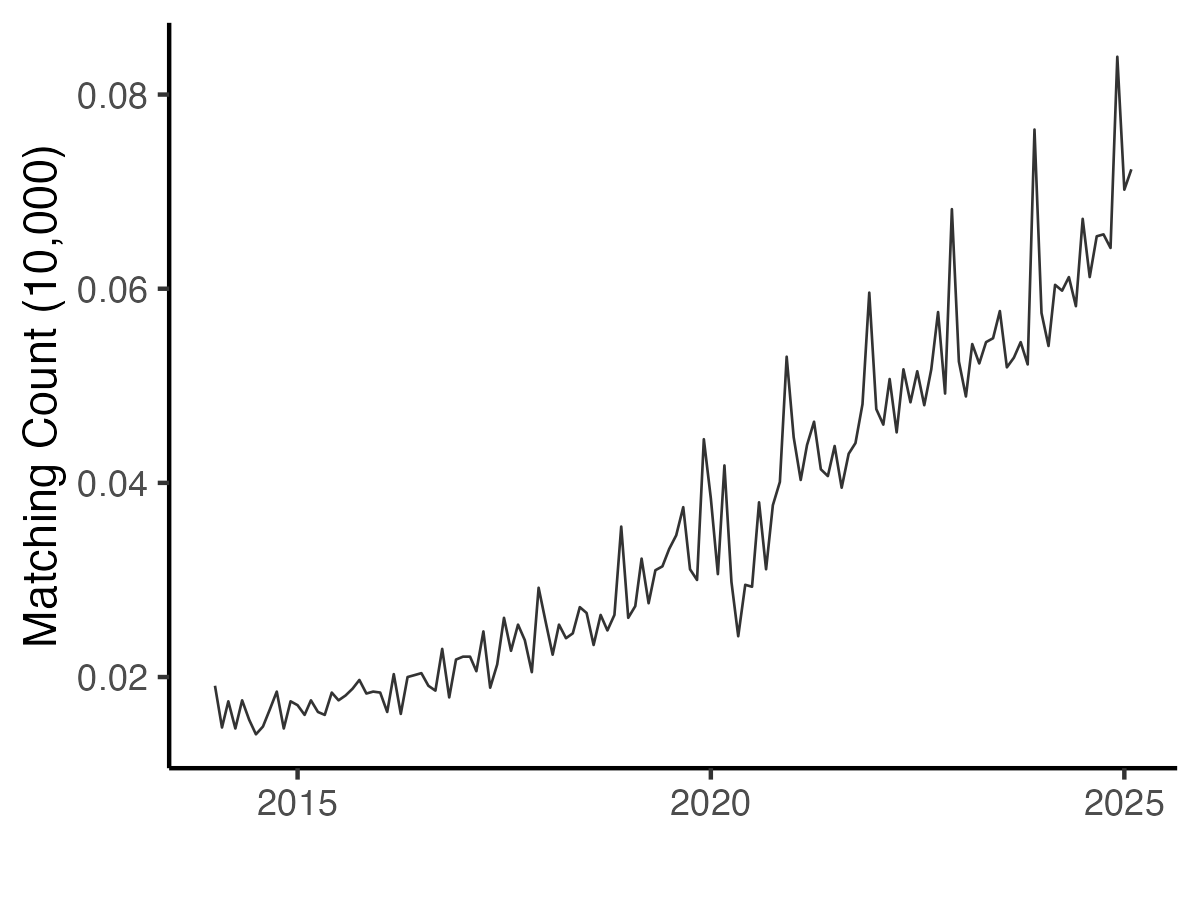}}
  \subfloat[Patner Finding Rate ($\frac{E}{F}$,$\frac{E}{M}$)]{\includegraphics[width = 0.33\textwidth]{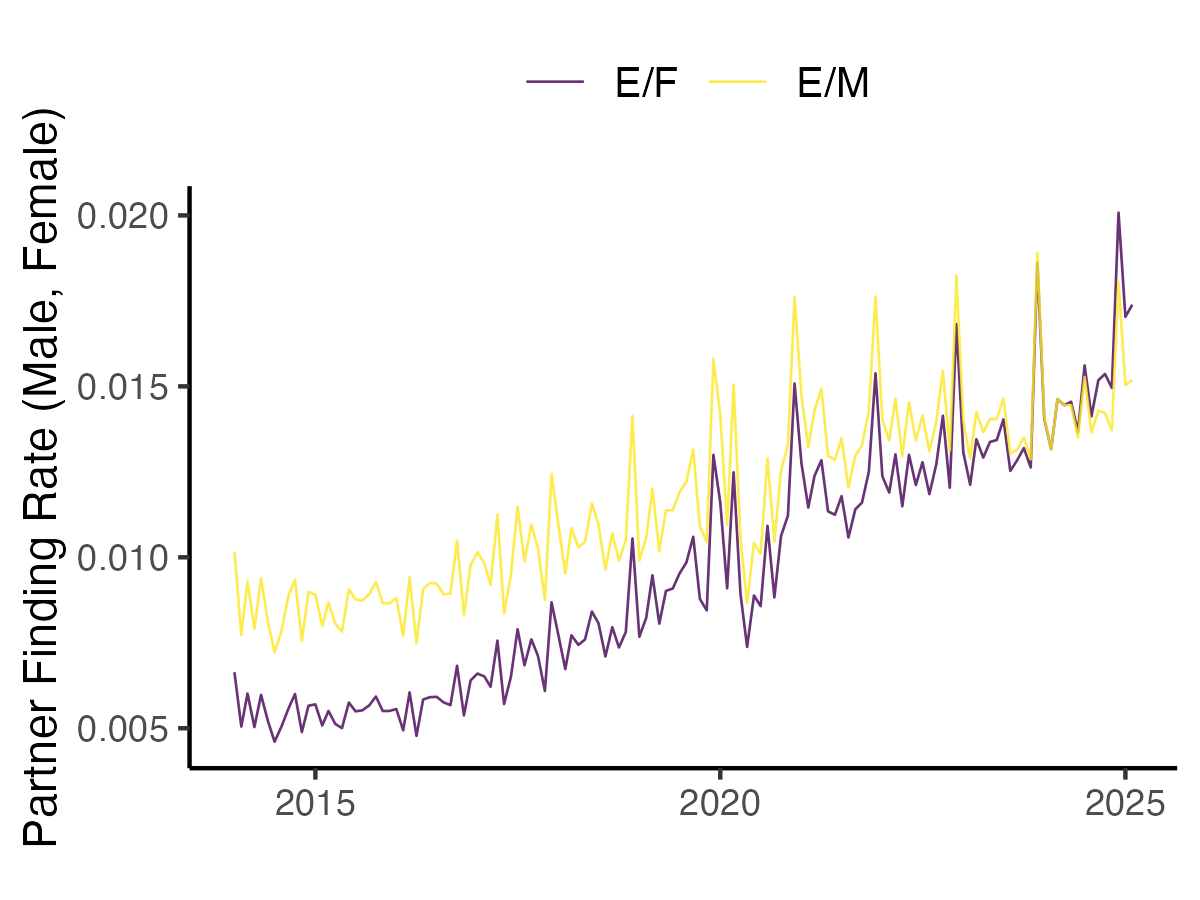}}
  \caption{Trends of key variables 2014-2025}
  \label{fg:country_month_male_female_partner_finding_rate} 
  \end{center}
  \footnotesize
  %Note: 
\end{figure}

Panels (a) and (b) in Figure \ref{fg:country_month_male_female_partner_finding_rate} display trends in the number of active users and engagements on the IBJ platform from 2014 to 2025. Both female $F$ and male $M$ participation rise steadily, with male user growth outpacing that of females—particularly after 2020—leading to a gradual decline in market tightness, measured as the gender balance ratio $ \frac{F}{M} $. Alongside this compositional shift, the number of engagements $E$, realized within the platform, remains relatively flat until around 2020, after which it increases sharply. This acceleration likely reflects improvements in platform performance, such as better algorithmic matching, increased activity, or broader adoption of digital matching services.

Panel (c) illustrates partner finding rates for female and male users per month, measured respectively as $ \frac{E}{F} $ and $ \frac{E}{M} $. Both rates increase over time, with females consistently achieving higher matching rates. This suggests improved match formation efficiency, particularly for female users, despite the growing supply of male participants. The male rate displays greater month-to-month volatility, indicating more sensitivity to market conditions.

\section{Empirical Framework}

This paper conceptualizes the marriage market as a two-sided search environment, where female users seek partners and male users are potential partners.\footnote{While an influential strand of the literature models marriage markets as stable, frictionless matching environments and estimates preference-based matching patterns using equilibrium assumptions (e.g., \citealp{chiappori2023mating}), this study takes a different empirical approach. Rather than recovering structural preferences under the assumption of stable matching, I focus on the empirical measurement of platform-level matching efficiency and elasticity over time. Our interest lies in understanding how the IBJ platform transforms search input into actual matches.} A successful match occurs when both parties agree to engage. Drawing on the matching function framework from labor economics, I estimate the number of engagements as a function of efficiency input and market tightness captured by the gender balance ratio.

Let $ E_t $ denote the number of within-platform engagements at time $ t $, $ F_t $ and $ M_t $ the number of active female and male users, and $ A_t $ a time-varying matching efficiency parameter. Efficiency $A_t$ measures how effectively the marriage market turns available men and women into engagements—for a given number of men and women, higher efficiency means more couples are formed. The matching efficiency can be arbitrarily interpreted as search effort, app-based improvements in user interface and experience, and so on (on the female side). Let $(A, F, M)$ denote random variables corresponding with realizations subscripted by time $t$. Let assume that $M$ and $A$ are independent conditional on $F$, that is $ M\mathop{\perp} A| F$. I assume a matching function of the form $ E_t = m(A_t F_t, M_t) $ with constant returns to scale, which is typically assumed in the literature. 
Given the assumptions, by applying the nonparametric identification results of \citet{matzkin2003nonparametric} to avoid functional form assumptions, Proposition 1 of \citet{lange2020beyond} shows that the observed joint distribution of engagements, female users, and male users, denoted as $ G(E, F, M) $, nonparametrically identifies the joint distribution of matching efficiency and female participation, $ H(A, F) $, as well as the matching function $ m(AF, M): \mathbb{R}^{2}_{+} \rightarrow \mathbb{R}_{+} $, up to a normalization of the efficiency parameter $ A $ at a given point $ A_0 $ in the support of $ (A, F, M) $.

\section{Results}

Using nonparametric methods proposed by \cite{lange2020beyond}, I estimate time-varying matching efficiency $ A_t $ and elasticities with respect to effective female search input $ A_t F_t $ and male users $ M_t $.\footnote{Details are shown in Appendix.} 

\subsection{Matching efficiency and elasticity in the marriage platform}

\begin{figure}[!ht]
  \begin{center}
  \subfloat[Matching Efficiency ($A$)]{\includegraphics[width = 0.45\textwidth]{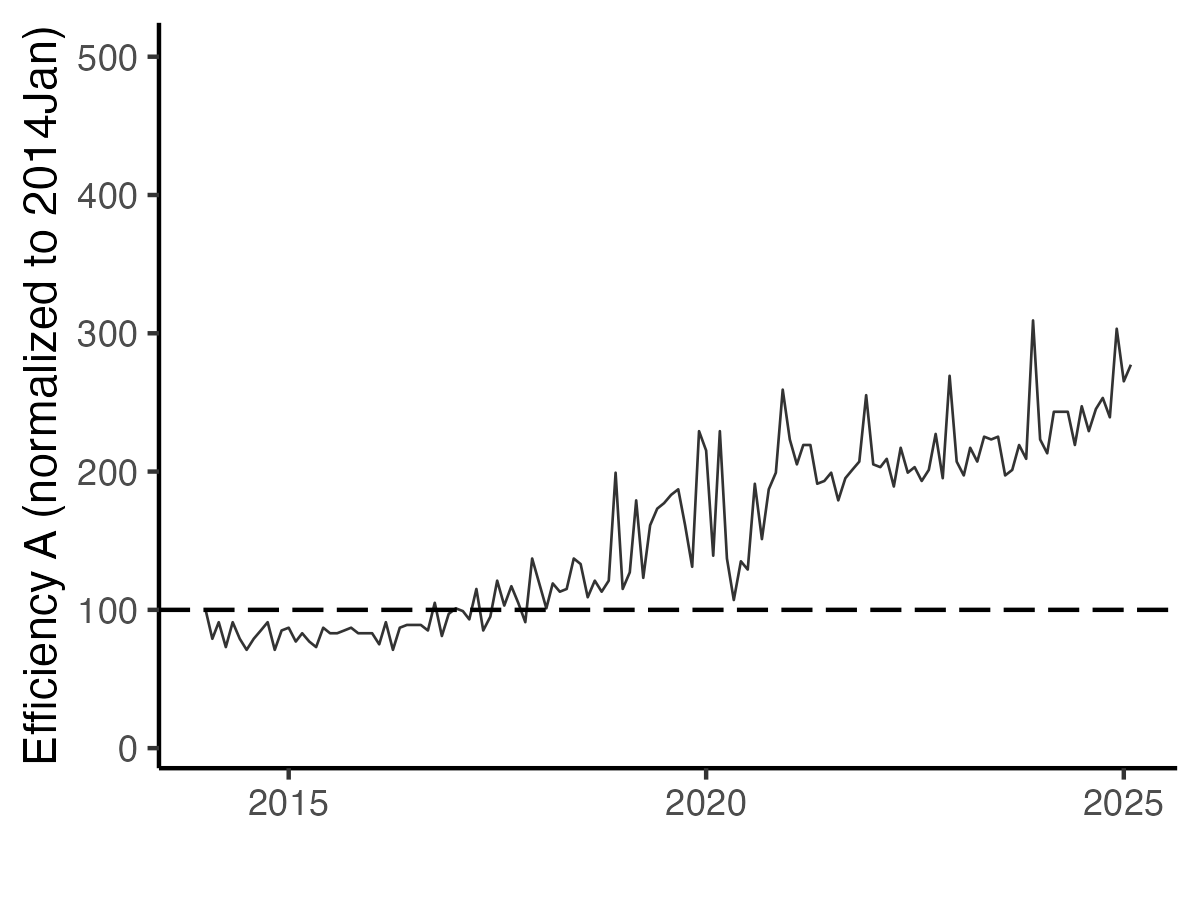}}
  \subfloat[Matching Elasticity ($\frac{d\ln m}{d \ln AF}$, $\frac{d\ln m}{d\ln M}$)]{\includegraphics[width = 0.45\textwidth]{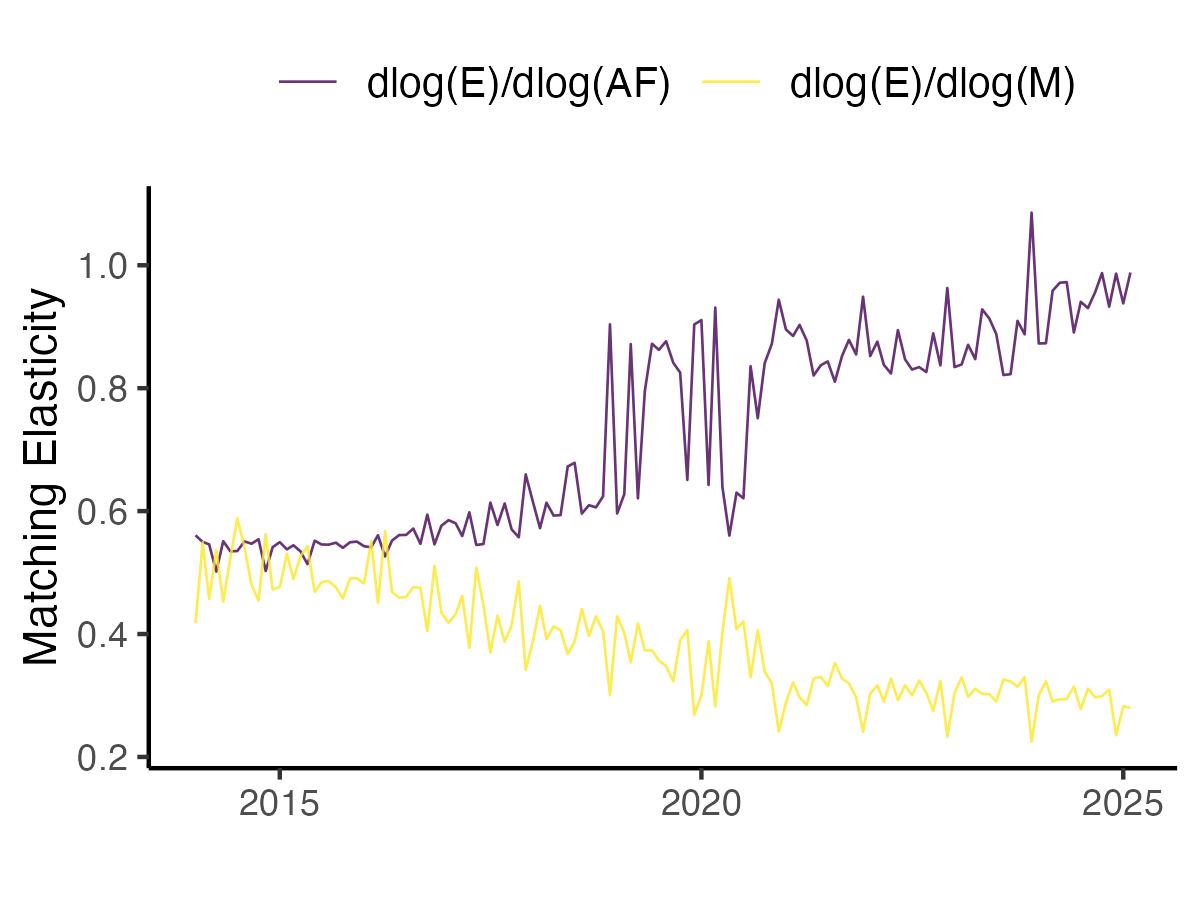}}
  \caption{IBJ platform 2014-2025}
  \label{fg:matching_efficiency_month_aggregate} 
  \end{center}
  \footnotesize
  Note: I normalize matching efficiency in January 2014 to 100.
\end{figure}

In Figure \ref{fg:matching_efficiency_month_aggregate}, matching efficiency increases threefold from baseline levels in 2014, with gradual acceleration beginning around 2017 and continuing through the post-2020 period. Notably, efficiency nearly triples between 2017 and 2025. This improvement may be partly attributable to structural changes within the platform, including the consolidation of other major marriage agencies into the IBJ network. For example, the integration of ZWEI—another nationwide marriage consultation service—into the IBJ Group in 2020 likely expanded the effective user base and improved cross-platform coordination.\footnote{Source: \textit{Nikkei (2020), “IBJ to Acquire Zwei to Expand Marriage Services,” March 11.}} Such developments may have enhanced matching opportunities and the sophistication of recommendation mechanisms, contributing to substantial gains in match formation efficiency.

Next, we calculate matching elasticity, which captures how sensitive the number of new couples is to changes in the number of available men or women. Elasticity is estimated by a lasso regression projecting matches on fully interacted 2nd order polynomials of $AF$ and $M$.
Elasticity with respect to female-side input ranges from 0.5 to 0.9, while male-side elasticity ranges from 0.5 to 0.3, reflecting an increasingly asymmetric impact of each side on overall matching outcomes.

Together, these trends indicate that the IBJ platform has become not only more efficient in producing engagements, but also more responsive—albeit asymmetrically—to changes in the user base composition. The rise in efficiency and divergence in elasticities suggest improved functionality in the underlying matching process, potentially driven by better recommendation algorithms, a more diverse user base, or targeted engagement strategies differentiated by gender.

\subsection{Regional differences in matching efficiency and elasticity}

\begin{figure}[!ht]
  \begin{center}
  \subfloat[Male ($M$)]{\includegraphics[width = 0.45\textwidth]{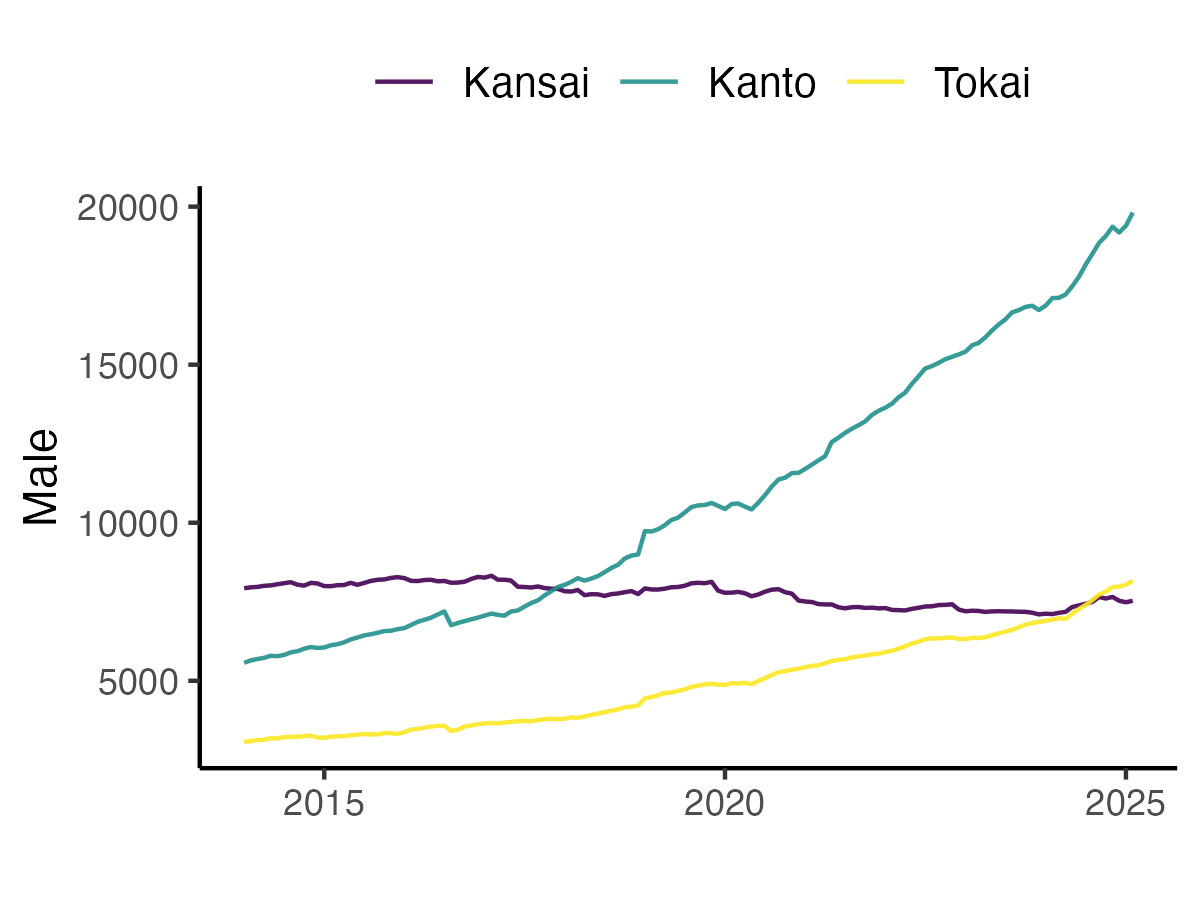}}
  \subfloat[Female ($F$)]{\includegraphics[width = 0.45\textwidth]{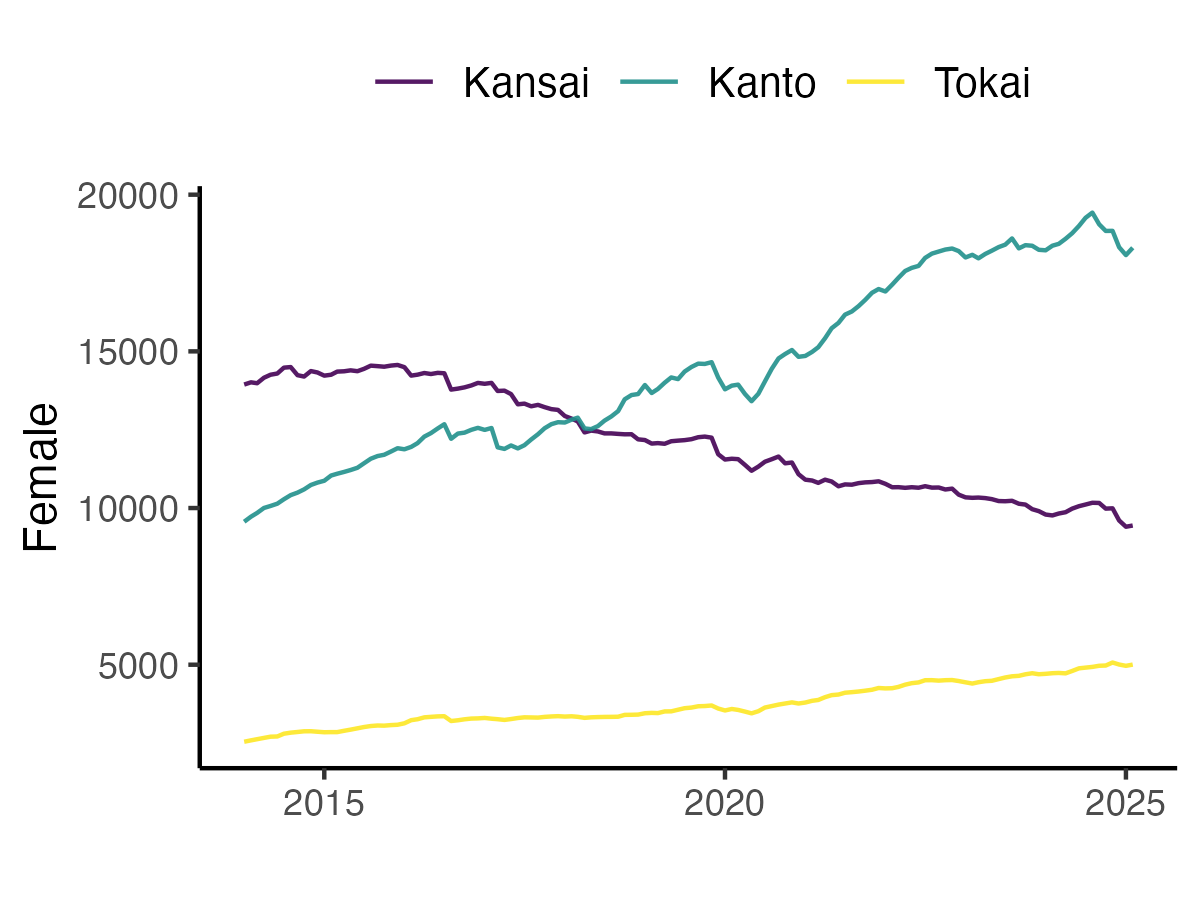}}\\
  \subfloat[Engage ($E$)]{\includegraphics[width = 0.45\textwidth]{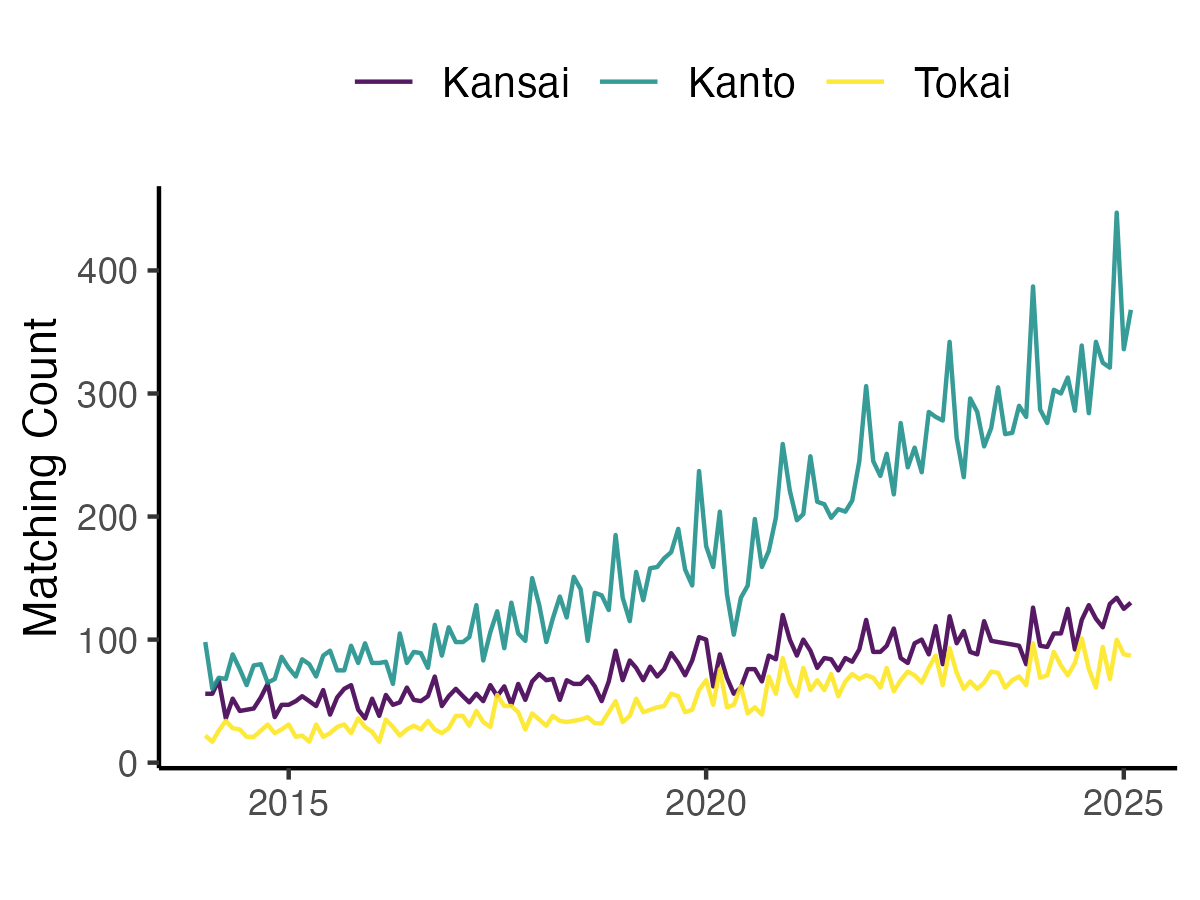}}
  %\subfloat[($V/U$)]{\includegraphics[width = 0.45\textwidth]  {figuretable/matching_function_project/tightness_month_aggregate_platform_each_industry.png}}\\
  \subfloat[Matching Efficiency ($A$)]{\includegraphics[width = 0.45\textwidth]{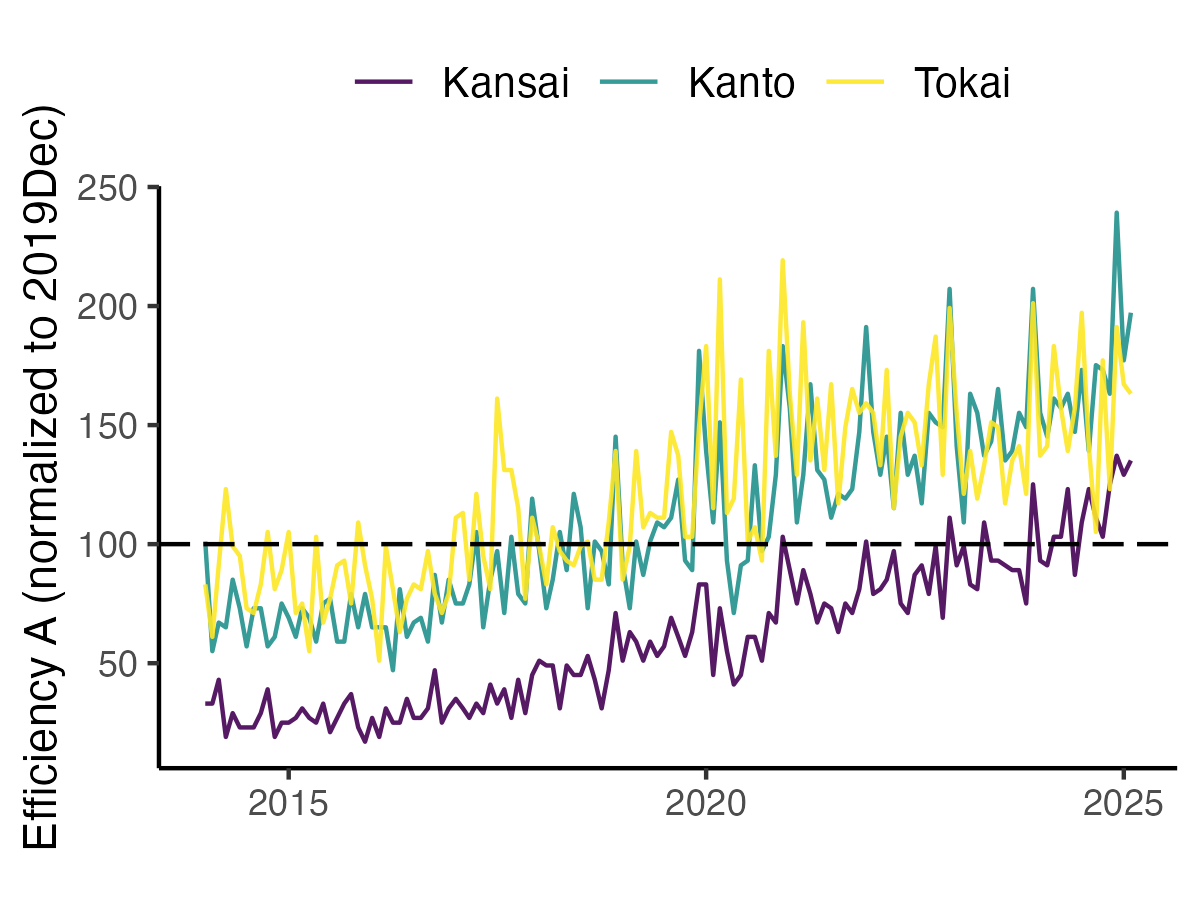}}\\
  \subfloat[Matching Elasticity ($\frac{d\ln m}{d \ln AF}$)]{\includegraphics[width = 0.45\textwidth]{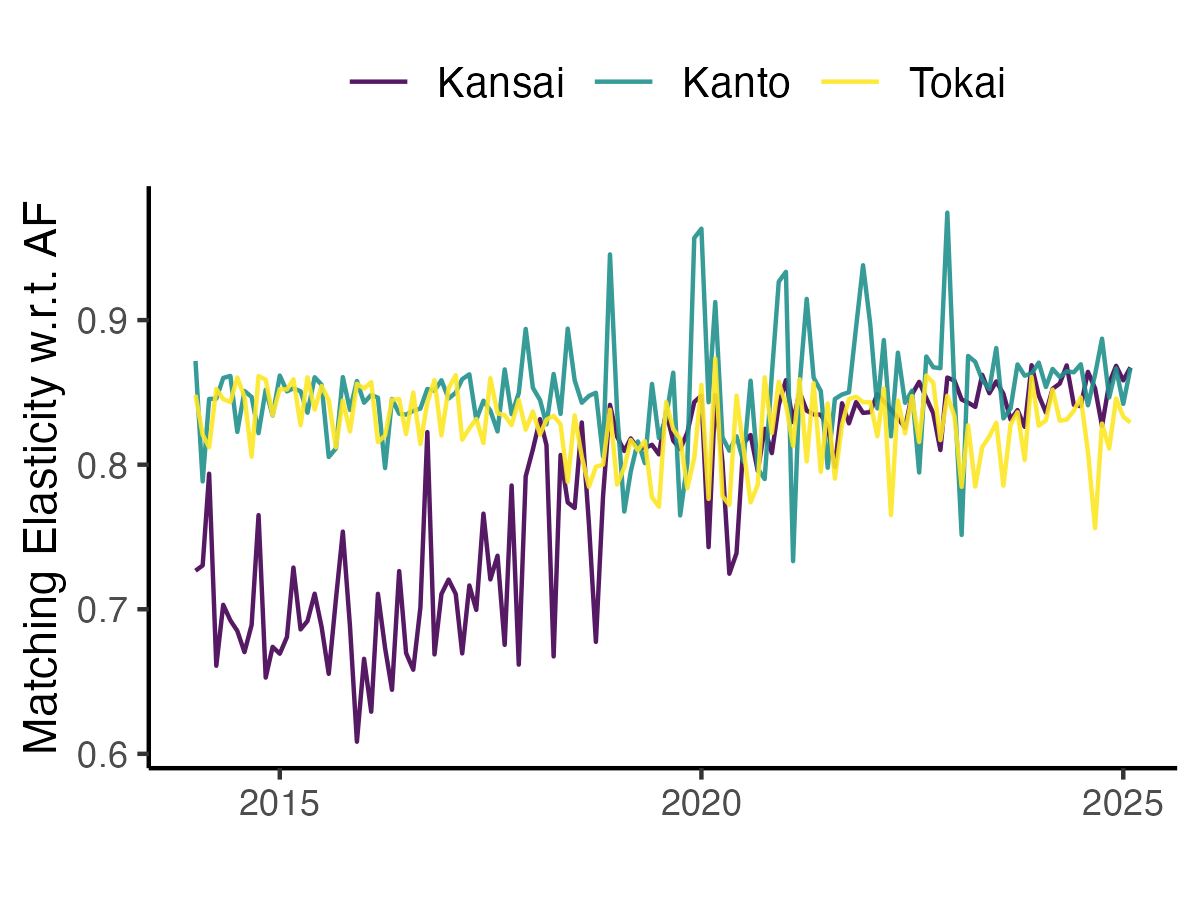}}
  \subfloat[Matching Elasticity ($\frac{d\ln m}{d\ln M}$)]{\includegraphics[width = 0.45\textwidth]{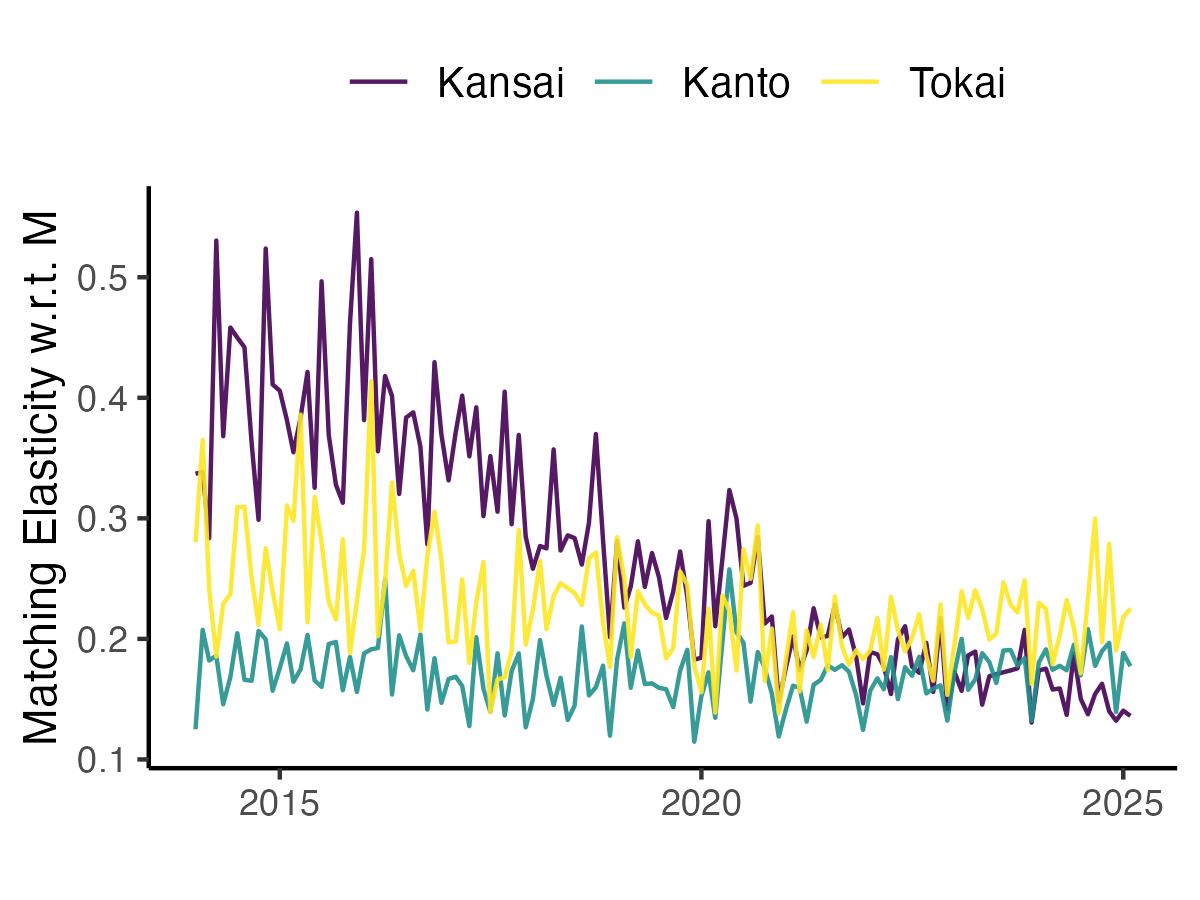}}
  \caption{IBJ platform 2014-2025}
  \label{fg:matching_efficiency_month_aggregate_each_industry} 
  \end{center}
  \footnotesize
  \footnotesize
  Note: I normalize matching efficiency in January 2014, Tokyo to 100.
\end{figure} 

I also explore regional variation across three major areas in Japan: Kanto (Tokyo), Kansai (Osaka), and Tokai (Aichi) in Figure \ref{fg:matching_efficiency_month_aggregate_each_industry}. All regions show improved matching efficiency normalized to January 2014, Kanto, with Kanto consistently leading. Elasticities are broadly similar across regions, suggesting stable platform functionality nationwide. Overall, the IBJ platform exhibits increasing matching efficiency and stable elasticity patterns across major metropolitan areas, reflecting both platform-level improvements and region-specific user growth.

\section{Conclusion}

This paper documents substantial improvements in matching efficiency on a large-scale marriage platform in Japan. Leveraging verified user data and well-defined engagement outcomes, I apply a nonparametric approach to estimate time-varying matching functions. The estimates show that the platform has become more effective and increasingly asymmetric in its responsiveness to user-side dynamics, with greater sensitivity to changes on the female side. Regional analysis confirms consistent functionality across major urban areas. These findings highlight the potential of digital matching platforms to reshape traditional partner search markets through structured intermediation and algorithmic improvements tailored to different sides of the market.

\newpage
\bibliographystyle{ecca}
\bibliography{ibj_project}

\newpage
\appendix
\section{Online Appendix (Not for publication)}
\subsection{Estimation details}
I begin by estimating the distribution function $ F(A_0 \mid F) $, following the identification strategy in \cite{lange2020beyond}. The logic hinges on the conditional distribution of engagements $ E $, given female and male user counts $ (F, M) $. Formally:

\[
F(A_0|\psi F_0) = G_{E|F,M}(\psi E_0|\psi F_0, \psi M_0),
\]

\[
F(\psi A_0|\lambda F_0) = G_{E|F,M}(\psi E_0|\lambda F_0, \psi M_0),
\]

where $ \psi $ is an arbitrary scalar, and $ \lambda $ is a scaling factor. By varying $ (\psi, \lambda) $, I trace out $ F(A \mid F) $ over the support of $ (A, F) $.

In practice, with finite data, I construct a nonparametric estimator of $ G_{E|F,M} $. For any evaluation point $ (E_\tau, F_\tau, M_\tau) $, I compute the proportion of observations with fewer engagements than $ E_\tau $ among those located near $ (F_\tau, M_\tau) $ in the $ (F, M) $-space. Kernel weights discount the influence of distant points. The estimator is written as:

\[
F(\psi A_0 | \lambda F_0) = G_{E|F,M}(\psi E_0 | \lambda F_0, \psi M_0),
\]

\[
\hat{F}(\psi A_0 | \lambda F_0) = \sum 1(E_t < \psi E_0) \cdot \kappa(F_t, M_t; \lambda F_0, \psi M_0),
\]

where $ \kappa(\cdot) $ is a bivariate normal kernel function with bandwidth set to 0.75. 

Once the distribution $ F(A \mid F) $ is recovered, I invert it using observed engagements to retrieve the implied matching efficiency $ A_t $ for each time period:

\[
A_t = F^{-1}(G(E_t | F_t, M_t) \mid F_t).
\]

Using the recovered values of $ A_t $, I then invert the matching function:

\[
m(A_t F_t, M_t) = G^{-1}(F(A_t \mid F_t) \mid F_t).
\]

Finally, I compute local elasticities of the matching function by 2nd order LASSO regression of engagements $ E $ on female and male user counts, interacted with the estimated effective search input $ AF $. This allows us to estimate the marginal responsiveness of engagements to changes in female and male participation:
\[
\text{Elasticity w.r.t. } F: \quad \frac{d \log m(AF, M)}{d \log F} = \frac{d \log m(AF, M)}{d \log AF},
\]
which is obtained by estimating the derivative of the matching function with respect to $ AF $, multiplied by the ratio $ \frac{AF}{E} $, using the regression coefficient from the empirical specification.

\subsection{Mobility across areas}

\begin{figure}[!ht]
  \begin{center}
  \subfloat[Main Land (Honshu) areas]{\includegraphics[width = 0.45\textwidth]{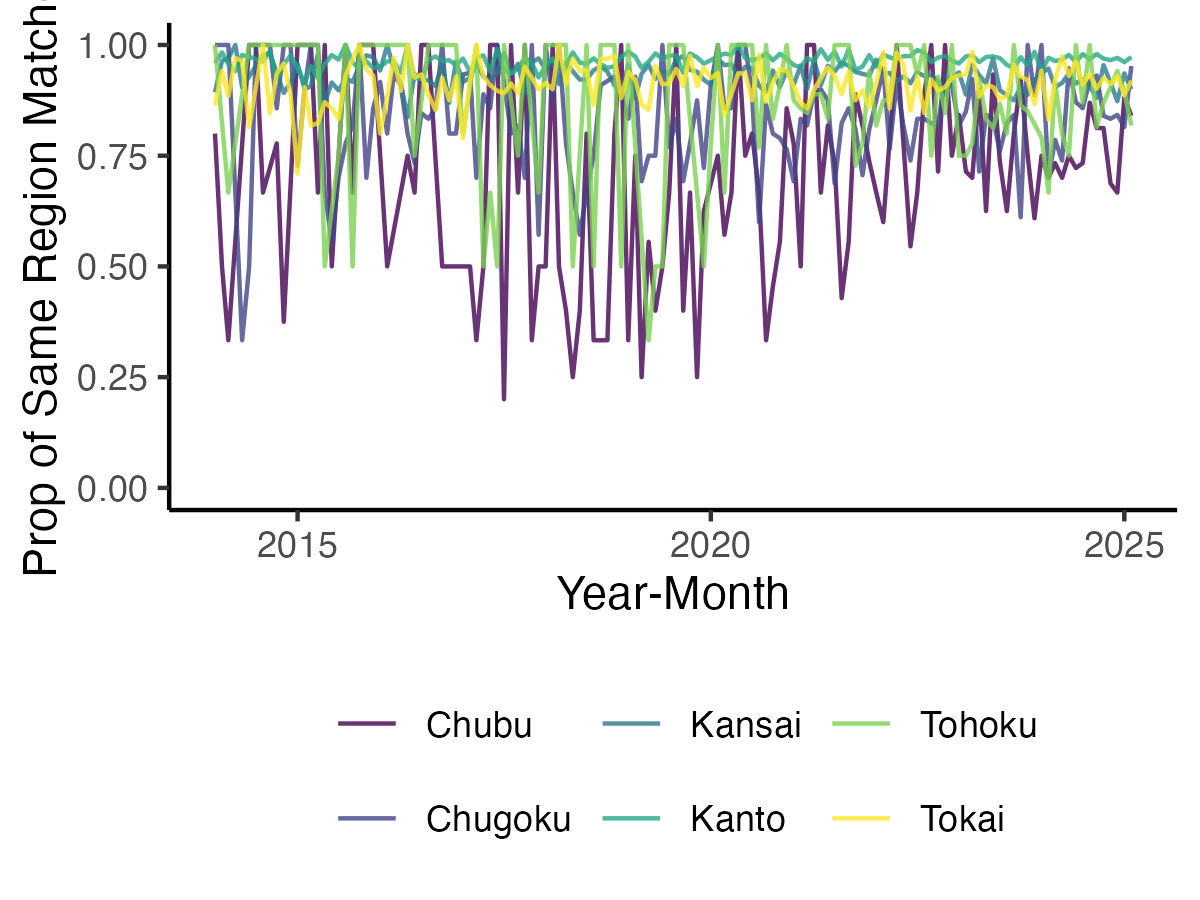}}
  \subfloat[Island areas]{\includegraphics[width = 0.45\textwidth]{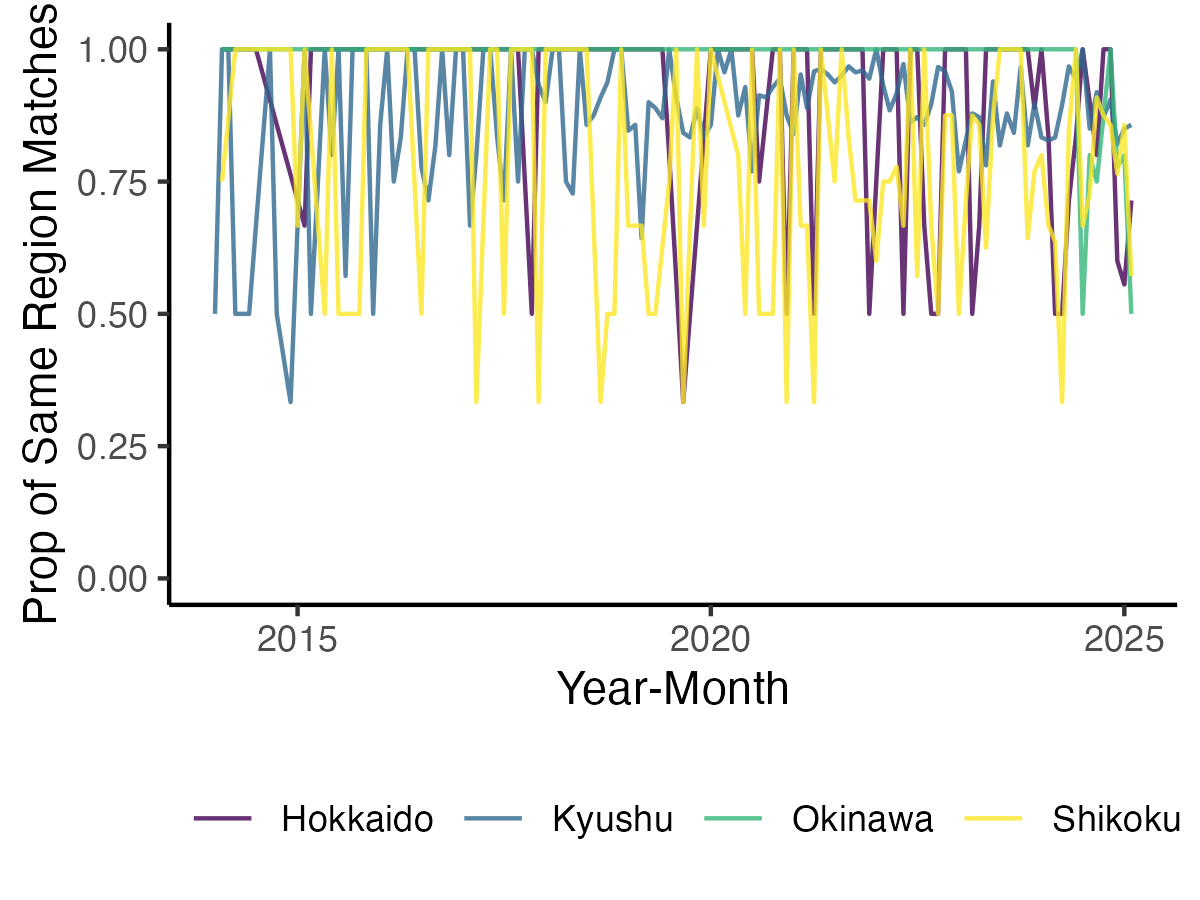}}
  \caption{Within-area Marriage Share}
  \label{fg:year_month_region_match_same_region_prop_island} 
  \end{center}
  \footnotesize
  %Note: For confidentiality reasons, I normalize the efficiency to 2019 December for each platform.
\end{figure}

Figure \ref{fg:year_month_region_match_same_region_prop_island} documents the share of matches formed between users residing in the same geographic area (“within-area matches”) over time. Panel (a) shows trends for major regions on the Japanese mainland (Honshu), while Panel (b) displays patterns for island regions. Across all regions, the majority of engagements occur between users residing in the same region, suggesting limited cross-regional mobility in the partner search process. Mainland regions such as Kanto and Kansai consistently exhibit within-region matching rates above 70–80\%, while smaller or less dense areas like Chugoku and Tohoku display greater fluctuation, especially in the earlier years of the sample.

In contrast, island regions such as Okinawa and Hokkaido show both higher volatility and a larger share of cross-regional matches, particularly in the early years, potentially reflecting more constrained local partner pools. Over time, the within-region matching rates for island areas also tend to converge toward mainland levels, suggesting an increase in local matching capacity or more geographically filtered search behavior. These patterns indicate that despite the digital nature of the platform, physical geography continues to play a substantial role in shaping match outcomes.

\end{document}